# Electron scattering dependence of dendritic magnetic instability in superconducting MgB$_2$ films

Z. X. Ye,[a)] Qiang Li,[a)] Y. F. Hu,[a)] A. V. Pogrebnyakov,[b)] Y. Cui,[b)] X. X. Xi,[b), c)]

J. M. Redwing,[c)] and Qi Li[b)]

[a)]Materials Science Department, Brookhaven National Laboratory, Upton, New York 11973
[b)]Department of Physics, Penn State University, University Park, PA 16802
[c)]Department of Materials Science and Engineering, Penn State University, University Park, PA 16802

Magnetic instability in both ultra-pure and carbon-doped MgB$_2$ films is investigated by magneto-optical imaging, transport and bulk magnetization measurements. In the carbon-doped MgB$_2$ thin films, familiar dendritic flux-jump patterns were observed at low temperature as reported in previous experiments. In the ultra-pure MgB$_2$ thin film, however, a remarkably stable flux penetration was observed, clearly showing the classic behavior of the critical state model. Such different behavior indicates that the electron scattering ultimately controls the magnetic stability of the MgB$_2$ films.

Correspondence should be addressed to qiangli@bnl.gov

Potential application of superconducting $MgB_2$ in power transmission cables, magnets, motors must take the magnetic stability of the conductor into account. A critical issue is stability against possible flux jumps. This is an avalanche process where flux motion dissipates heat and leads to a local temperature rise which reduces local pinning and facilitates further flux motion.[1,2] Unprotected flux jumps can sometimes result in a thermal runaway and destroy superconducting equipment. Strong flux instabilities in thin films and bulk $MgB_2$ at low temperatures have been reported.[3-6] Using magneto-optical imaging (MOI) techniques, Johansen *et al.* revealed that below 10 K the penetration of magnetic flux in pulsed-laser-deposited (PLD) $MgB_2$ films was dominated by dendritic structures abruptly formed in response to an applied field.[3] Though dendritic magnetic instability is believed to be of thermomagnetic origin, the phenomenon is still poorly understood. Recently, Johansen *et al.*[3] and I. Aranson *et al.*[7] numerically simulated dendritic flux jump in type II superconductors based on the thermal feedback mechanism. However, these simulations are unable to relate the occurrence of flux jump quantitatively to the parameters specific to a particular superconductor, such as its critical temperature $T_c$, upper critical field $H_{c2}$, normal state resistivity $r_n$, and critical current density $J_c$, etc. Furthermore, the central issue of what controls the magnetic stability, particularly in $MgB_2$, remains unsolved.

The goal of the present work is to explore suppression of the magnetic instability in $MgB_2$ by varying materials properties, and to determine the key factor responsible for the dendritic flux jumps in $MgB_2$ films. Experiments focused on high quality $MgB_2$ thin films made by hybrid physical-chemical vapor deposition (HPCVD).[8] It has been shown previously that the HPCVD process is very successful in producing ultra-pure MgB$_2$ films with very low resistivity and clean-limit behavior ("ultra-pure"), being much cleaner than even pure films made by PLD or other route.[9] In addition, films with a broad range of "dirty-limit properties" could also be obtained through carbon doping via HPCVD process.[10] Magnetic behaviors of these films were studied by a combination of MOI, transport, and bulk magnetization measurements. These results were then be compared to our earlier assessments of pure MgB$_2$ films made by PLD.[5] Our studies showed that dendritic flux jumps can be completely eliminated by keeping $r_n$ sufficiently low, such as in the ultra-pure MgB$_2$ thin films made by HPCVD.

Two types of MgB$_2$ film (ultra-pure and C-doped) were grown on c-cut SiC single crystalline substrates using the *in-situ* HPCVD process described previously.[8] The ultra-pure films are clean and *epitaxial* with the c-axis perpendicular to the surface, while the C-doped films are uniaxially oriented with the c-axis perpendicular to the surface, similar to most of the PLD films reported so far. Films sized 5×5 mm$^2$ of ultra-pure MgB$_2$ (330 nm thick) and 0.12 formula carbon doped MgB$_2$ (200 nm thick) were selected for MOI study. $T_c$ are 41.2 K and 38.4 K for the ultra-pure and the C-doped samples, respectively. Flux motion in these MgB$_2$ thin films was directly recorded using the high resolution MOI station described elsewhere.[5] Bulk magnetization was measured on the *same* films in a Quantum Design MPMS SQUID magnetometer with the applied field $H_a \perp$ film surface.

Fig. 1 shows the MO images. Striking differences in the flux penetration patterns are immediately apparent between C-doped (a and b) and ultra-pure MgB$_2$ films (c and d). These images were taken at 4.2 K during the initial magnetization (a and c) after zero-field-cooling and in the remanent states (b and d). Fig. 1a shows vigorous dendritic flux penetration into the C-doped film at external field $\mu_o H_a$ = 8 mT. The dendrites nucleate randomly near the film edge and propagate into the films immediately. Fig. 1b shows the remanent state of the C-doped film after $\mu_o H_a$ was reduced from 0.1 T to zero, where the dark dendrites show the sudden exit of flux. In general, the behavior of the dendritic flux motion in these C-doped films is essentially the same as that observed previously in pure MgB$_2$ films made by PLD.[3,5] Similarly, at $T > 10$ K, MOI did not reveal any flux jumps in the C-doped films. In a striking contrast, dendritic flux jumps are completely absent in the ultra-pure MgB$_2$ films. Instead, a regular and gradual flux penetration was observed, as shown in Fig. 1c taken at $\mu_o H_a$ = 20 mT and in Fig. 1d taken after $\mu_o H_a$ was reduced from 0.1 T to zero, being consistent with the prediction of the critical state model. The MO images of Fig. 1c and 1d are not indicative of completely homogenous flux penetration, as can be seen in some other superconducting systems.

The magnetic behavior of the C-doped and the ultra- pure MgB$_2$ films was further explored by bulk magnetization measurements. Fig. 2 shows that the initial magnetization of the C-doped film contains small flux jumps, as indicated by pronounced noise illustrated in the inset to Fig 2. Such behavior was observed at 1.8 K $\leq T \leq$ 9.5 K for $\mu_o H_a$ up to 0.15 T. At $T \geq$ 10 K, both initial and full hysteresis measurements gave smooth magnetization curves for the C-doped film. In contrast, smooth magnetization curves were always observed at all temperatures, as low as 1.8 K, for the ultra-pure film. These observations are in excellent agreement with those flux profiles found in the MOI studies above.

To determine the key factors responsible for the disappearance of the dendritic flux jumps in the ultra-pure MgB$_2$ films, we analyze various parameters relevant to the thermomagnetic instability. We first examine $J_c(T)$. In general, the higher the value of $J_c$ is, the steeper the slope of the flux gradient in the critical state is, and hence the higher is the tendency for a breakdown of a critical state with respect to hot spot formation. In addition, higher $|dJ_c/dT|$ gives faster propagation of this breakdown. Fig. 3 shows the $T$-dependence of $J_c$ obtained for the C-doped and the ultra-pure films from both transport measurements of similarly processed films in self field,[11] and magnetization measurements in remnant field of the actual films viewed by MOI (the standard Bean model is applied to magnetic hysteresis). The large error bar in $J_c$ for the C-doped film at 4.2 K is due to slight differences in C-concentration among the films. In fact, the ultra-pure film has higher $J_c$ and $|dJ_c/dT|$ than the C-doped film. This rules out the possibility that $J_c$ is responsible for the absence of dendritic flux jumps in the ultra-pure MgB$_2$ films

Next, we examine the conditions required for local flux jumps, which depend on the ratio $\tau$ of the flux ($t_m$) and thermal ($t_t$) diffusion time constants.[7] The dimensionless parameter $\tau$ is given by $\tau = t_m / t_t = D_t / D_m = \mu_0 k / C\rho_f$, where the magnetic diffusivitiy $D_m = \rho_f/\mu_0$, thermal diffusivity $D_t = k/C$, $\mu_0$ is the permeability of vacuum; and $k$, $C$, and $\rho_f$ are the superconductor's thermal conductivity, heat capacity, and flux flow resistivity, respectively. Under local adiabatic conditions, where $\tau \ll 1$ ($t_m \ll t_t$), the magnetic flux diffusion is considerably faster than that of heat, and there is not enough time to redistribute and remove the heat released due to flux motion. Under these conditions, dendritic flux jumps occur. Such conditions applied to earlier experiments on thin films of Nb[12] and MgB$_2$ as discussed earlier.[3, 5] Although it is not feasible for us to measure the thermal conductivity and the heat capacity of these MgB$_2$ films directly, other works on the C-doped MgB$_2$ samples suggest that these parameters should not be too different from those of pure MgB$_2$.[13] It becomes clear that the key parameter that changes the local adiabatic condition is the flux flow resistivity $r_f$. For simplicity, $r_f$ is approximated by $r_f = r_n H_a/H_{c2}$.[14] It is clear that $r_f$ must be greatly reduced in order to slow down the flux diffusion. This can be accomplished by either a drastic reduction of $r_n$ or a large increase of $H_{c2}$. In one-band superconductors these trends oppose each other, $H_{c2}$ being proportional to $r_n$, and thus it is not possible to separately alter $r_n$ and $H_{c2}$ easily.[14] However, in two-band superconductors it is possible to embody strong electron scattering and high $H_{c2}$ at low temperature in one dirty band while keeping low $r_n$ in the other clean band.[15]

Fig. 4 shows the $T$-dependence of resistivity for the ultra-pure and C-doped MgB$_2$ films at zero field, plotted along with the result for the PLD film reported earlier.[5, 16] $T_c$ is similar for all three films (~ 38 – 41 K), while the difference in $r_n$ at $T_c$ is huge (~ 0.4 μΩ·cm for the ultra-pure and ~ 40 μΩ·cm for the C-doped film, respectively, while for the earlier PLD film, $r_n$ at $T_c$ is ~ 7 μΩ·cm). This dramatic change in resistivity (over two orders of magnitude) is not compensated for by three fold changes in $H_{c2}$. $\mu_0 H_{c2}$ for the C-doped film was found to be ~ 25 T for $H$ //c at 4.2 K, based on the C-doping, as compared to 7 T in the ultra-pure MgB$_2$ films.[10] Taking $H_a$ = 0.1 T as an example, $r_f$ for the ultra-pure MgB$_2$ film is ~ 0.006 μΩ·cm, versus 0.16 μΩ·cm in our carbon doped sample. For the earlier pure MgB$_2$ film made by PLD,[5, 16] $r_f$ is ~ 0.1 μΩ·cm, where $H_{c2}$ ~ 7 T at 4.2 K was used. In fact, our C-doped film and earlier PLD films have comparable $r_f$ at low temperatures. Thus, it is not surprising to observe similar dendritic flux jumps in those films. In comparison, $r_f$ in the ultra-pure $MgB_2$ films is about two orders of magnitude lower. This drastically reduced $r_f$ in the ultra pure $MgB_2$ film slows flux diffusion considerably, and hence prevents dendritic flux jumps.

In summary, we report the absence of dendritic flux jumps in the ultra-pure $MgB_2$ films at temperatures as low as 1.8 K for the first time. Similar dendritic flux jump behavior observed in C-doped $MgB_2$ films and earlier pure $MgB_2$ films made by PLD is likely due to their comparable values of flux flow resistivity. It is shown that the magnetic stability in $MgB_2$ films is closely related to their normal state resistivity. The onset of dendritic instabilities in dirty-limit $MgB_2$ films perhaps manifests another unique feature of electron scattering in the two-band superconductors.

ACKNOWLEDGMENT

The authors would like to thank Dr L. Cooley for critical reading of this manuscript. The work at BNL was supported by the U. S. Dept. of Energy, Office of Basic Energy Science, under contract No. DE-AC-02-98CH10886. The work at Penn State is supported in part by ONR under grant Nos. N00014-00-1-0294 (Xi) and N0014-01-1-0006 (Redwing), by NSF under grant Nos. DMR-0306746 (Xi and Redwing), DMR-0405502 (Qi Li).

FIGURE CAPTIONS

Fig. 1  Magneto-optical images showing flux penetration patterns in the zero-field-cooled C-doped and ultra-pure MgB$_2$ films (5×5 mm$^2$) at 4.2 K. (a) Dendritic flux penetration into the C-doped film at $\mu_o H_a$ = 8 mT. (b) Remanent state of the C-doped film showing the exit of the dendritic flux after $\mu_o H_a$ reduced from 0.1 T to zero. (c) Stable and gradual flux penetration into the ultra-pure film at $\mu_o H_a$ = 20 mT. Note: at $\mu_o H_a$ < 10 mT, nearly perfect shielding (no flux entry) was observed. (d) Remanent state of the ultra-pure film showing regular roof-top flux trapping pattern after $\mu_o H_a$ reduced from 0.1 T to zero.

Fig. 2  Initial magnetization of the C-doped and ultra-pure MgB$_2$ films. The inset is an expanded view of the magnetization curve for the C-doped film at 5 K, showing pronounced noise due to flux jumps. A regular magnetization behavior was observed for the ultra-pure MgB$_2$ film at all temperatures from 1.8 K to $T_c$, as well as for the C-doped film at $T \geq 10$ K.

Fig. 3  Temperature dependence of $J_c$ for the C-doped and ultra-pure MgB$_2$ films determined by the transport (Tran.) and magnetization (Mag.) methods

Fig. 4  Temperature dependence of resistivity of the C-doped and ultra-pure MgB$_2$ films plotted along with the pure MgB$_2$ film made by PLD (Ref. 5 and 16)

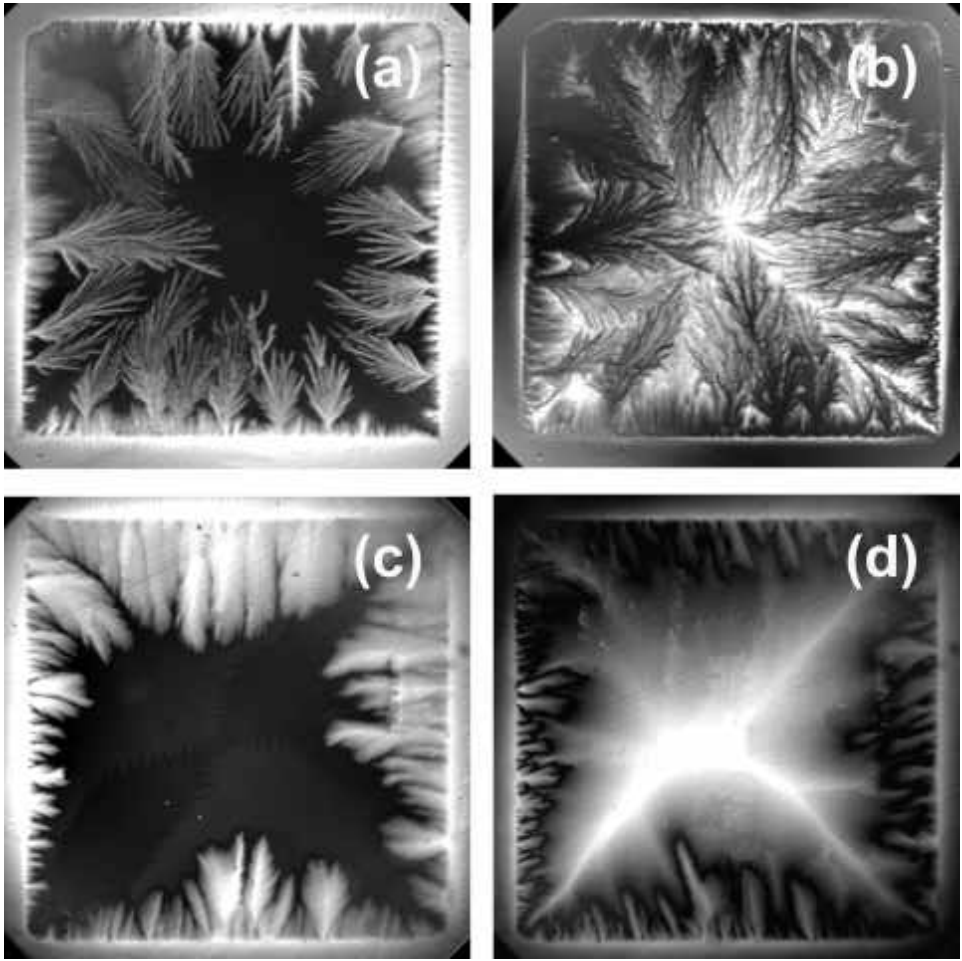

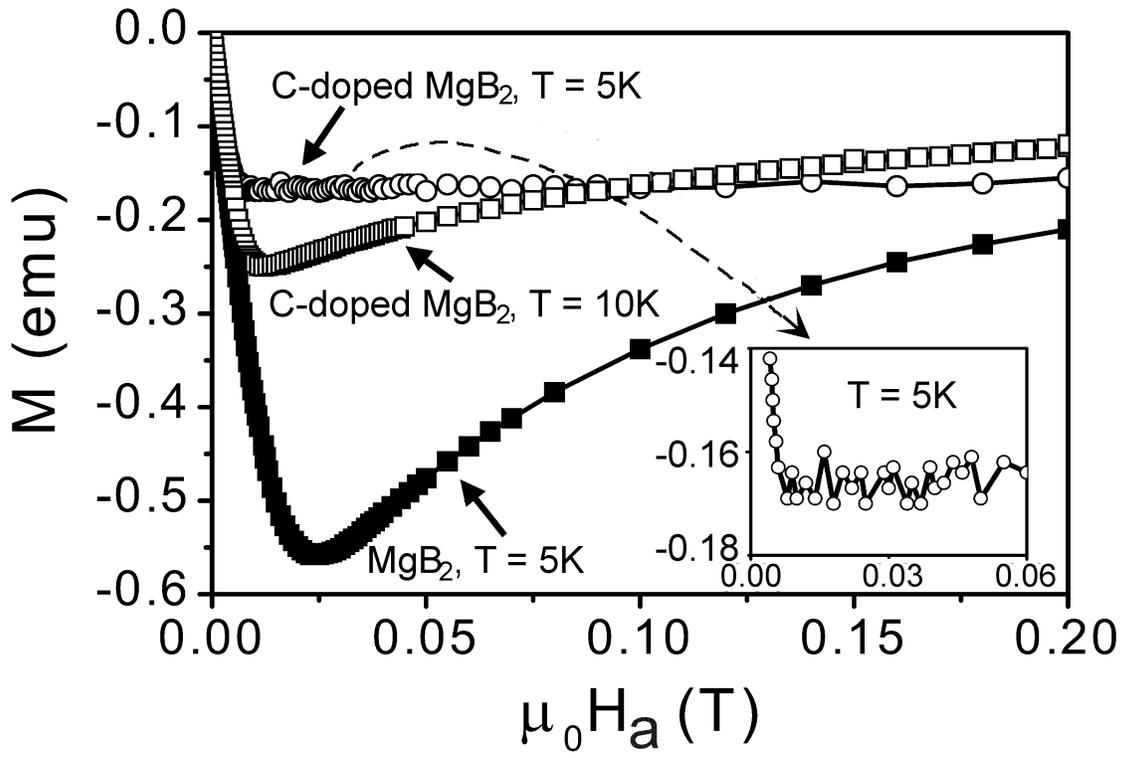

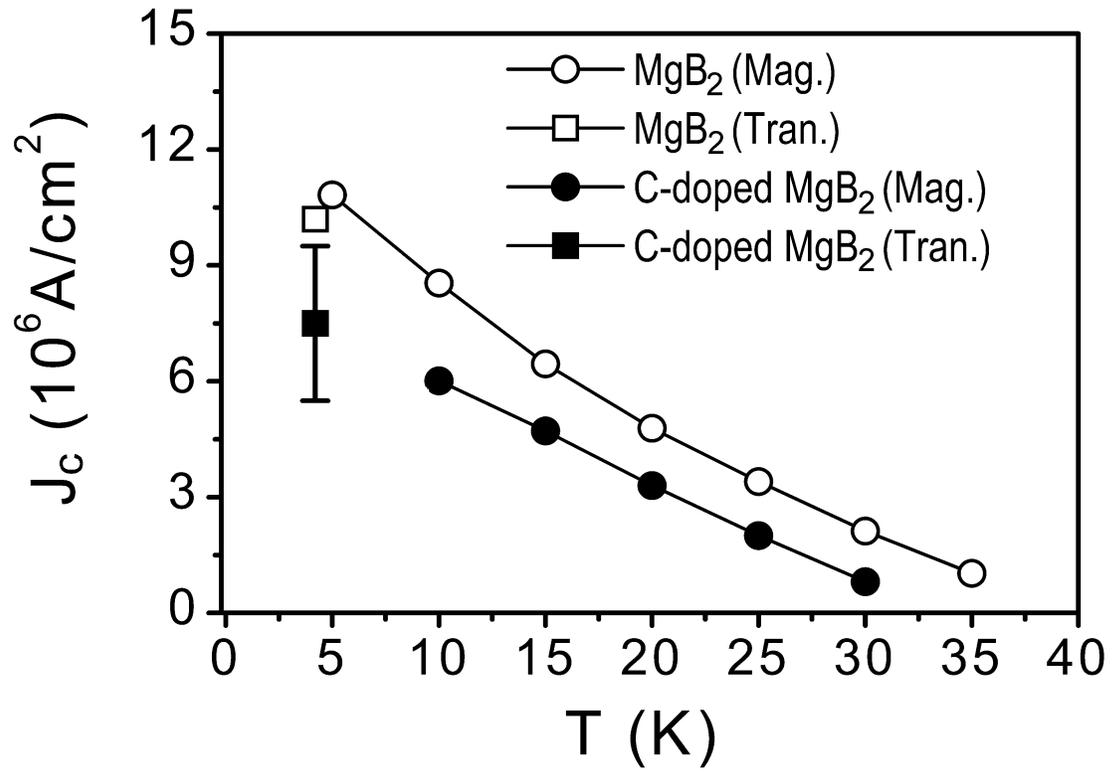

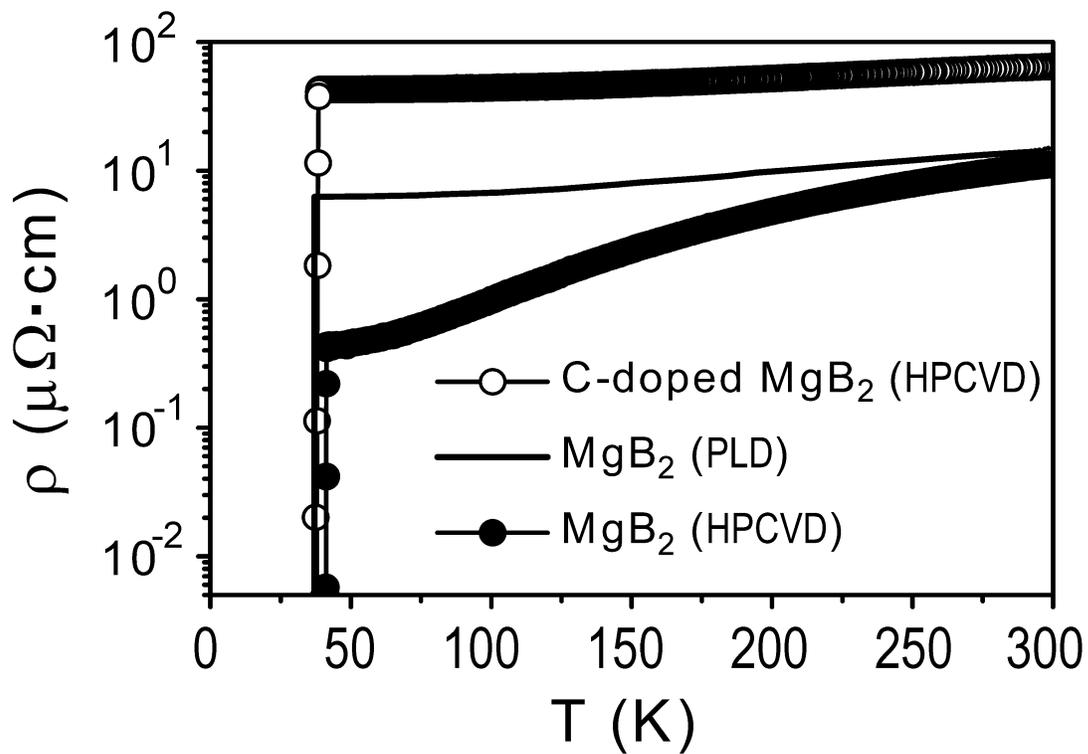